\documentclass[aps,prb,twocolumn,groupedaddress,showpacs,intlimits,amsmath,amssymb,floatfix]{revtex4-1}

\usepackage{bm}
\usepackage[final]{graphicx}
\usepackage[hang,center]{subfigure}
\usepackage{color}

\begin{document}

\title{Quasiparticle interference and the interplay between superconductivity and density wave order in the cuprates}

\author{E.A. Nowadnick$^{1,2}$}
\author{B. Moritz$^{2,3,4}$}
\author{T.P. Devereaux$^{2}$}
\affiliation{$^1$Department of Physics, Stanford University, Stanford, CA 94305, USA.}
\affiliation{$^2$Stanford Institute for Materials and Energy Science, SLAC National Accelerator Laboratory and 
Stanford University, Stanford, CA 94305, USA}
\affiliation{$^3$Department of Physics and Astrophysics, University of North Dakota, Grand Forks, ND 58202, USA}
\affiliation{$^4$Department of Physics, Northern Illinois University, DeKalb, IL 60115, USA}

\date{\today}

\begin{abstract}
Scanning tunneling spectroscopy (STS) is a useful probe for studying the cuprates in the superconducting and pseudogap states. Here we present a theoretical study of the Z-map, defined as the ratio of the local density of states at positive and negative bias energies, which frequently is used to analyze STS data.  We show how the evolution of the quasiparticle interference peaks in the Fourier transform Z-map can be understood by considering  different types of impurity scatterers, as well as particle-hole asymmetry in the underlying bandstructure. We also explore the effects of  density wave orders, and show that  the Fourier transform Z-map may be used to both detect and distinguish between them.
\end{abstract}

%
%
 \maketitle
\section{Introduction}

Conventional metals and superconductors  exhibit particle-hole symmetry at low energies, which is well described by Fermi liquid and BCS theory.  In contrast, a notable feature of the high-$T_c$ cuprate superconductors is the number of ways in which they break particle-hole symmetry.  Electron or hole doping the Mott insulating parent compounds leads to superconductivity, although the maximum $T_c$ is drastically different, reaching 150 K for hole doped compounds, while only 30 K in the case of electron doping. Another example is a pronounced asymmetry in the conductance spectrum for tunneling electrons and holes, measured by scanning tunneling spectroscopy (STS).~\cite{stm_review} These  asymmetries can be understood in the context of doping a Mott insulator, where the addition of electrons is more difficult than extraction, due to the strong electron-electron repulsion.~\cite{mohit, anderson_ong} 

Recent experiments have  observed further particle-hole symmetry breaking in the pseudogap state, a phase whose origin is debated, but may be due to preformed Cooper pairs lacking phase coherence, or a competing  order.~\cite{Norman}  Angle-resolved photoemission spectroscopy (ARPES)~\cite{arpes_review} in the pseudogap regime reveals particle-hole symmetry breaking in the form of a bend-back in the energy dispersion misaligned with the Fermi momentum ${\bf k}_F$, which can arise from a density wave (DW) order.~\cite{Makoto, ruihua}  Because superconducting fluctuations should be roughly particle-hole symmetric, while spin and charge orders generally induce particle-hole asymmetric changes, searching for particle-hole symmetry breaking is a useful way to distinguish between superconductivity and DW orders.

  In this paper we explore the signatures of particle-hole asymmetry in the local density of states (LDOS) measured by STS, a technique which has been used extensively to study the cuprates in both the superconducting and pseudogap phases.~\cite{stm_review,balatsky_stm_review,alloul}
 STS has revealed that the cuprates have a spatially inhomogeneous electronic structure, including modulations in the LDOS and superconducting gap magnitude.~\cite{howald_2001,pan,Hoffman, hoffman_cb,lang, McElroy_qpi1,kohsaka_2004, hanaguri_cb, Vershinin, McElroy_qpi2,  Hanaguri_qpi, Kohsaka_glass, Kohsaka_How_CP_vanish,Wise_cdw,Wise,lee_stm,Parker, davis_review}    In the $d$-wave superconducting phase, the LDOS modulations can arise from quasiparticle interference (QPI), due to the scattering of wave-like quasiparticles off impurities.~\cite{Hoffman, lee_stm, McElroy_qpi1,Kohsaka_How_CP_vanish, davis_review, Parker}   The wavevectors of the modulations can be determined from the Fourier transform of the LDOS.

Fourier transform STS (FT-STS) data are often analyzed using the Z-map, defined as the ratio of the LDOS at positive and negative bias energies,
\begin{equation}
\label{eq:z}
Z({\bf r},\omega)=\frac{n({\bf r},+\omega)}{n({\bf r},-\omega)}.
\end{equation}
 Experimentally, the Z-map has been used to cancel both non-dispersing ``checkerboard" modulations and systematic errors due to tip elevation uncertainty, in order to reveal the superconducting QPI pattern.~\cite{Kohsaka_How_CP_vanish, Hanaguri_qpi} By definition, the magnitude of the Z-map clearly carries information on particle-hole asymmetry in the LDOS.  
 
 Due to superconducting coherence factors, the Z-map in fact enhances the intensity of LDOS modulations from QPI,  as discussed in Ref.~\onlinecite{fujita_ba}.  To summarize their argument, since STS tunnels electrons rather than quasiparticles, the strength of the measured conductance at  ${\bf r}$ depends on the magnitude of the hole and electron amplitudes, $|u_n({\bf r})|^2$ and $|v_n({\bf r})|^2$ ($n$ labels the eigenvalue of excitations).  Expressing the LDOS in terms of the electron Green's function $n({\bf r},\omega) = -(1/\pi) Im \hat{G}_{11}({\bf r},\omega)$, the dependence of the LDOS on coherence factors is clear:
 \begin{eqnarray}
 n({\bf r},\omega>0) &=& \sum_n |u_n({\bf r})|^2 \delta(\omega-E_n) \\ \nonumber
 n({\bf r},\omega<0) &=& \sum_n |v_n({\bf r})|^2 \delta(\omega+E_n).
 \end{eqnarray}
 Assuming well defined excitations in energy (so the sum over $n$ includes just one term), the Z-map  reduces to
 \begin{equation}
 Z({\bf r},\omega)=\frac{|u({\bf r})|^2}{|v({\bf r})|^2}
 \end{equation}
 By recalling the sum rule $\sum_n |u_n({\bf r})|^2+|v_n({\bf r})|^2=1$, it is clear that whenever $|u({\bf r})|^2$ is large $|v({\bf r})|^2$ is small, so taking the ratio of the coherence factors will enhance the maxima.  
 
\begin{figure}

\includegraphics[width=0.48\textwidth]{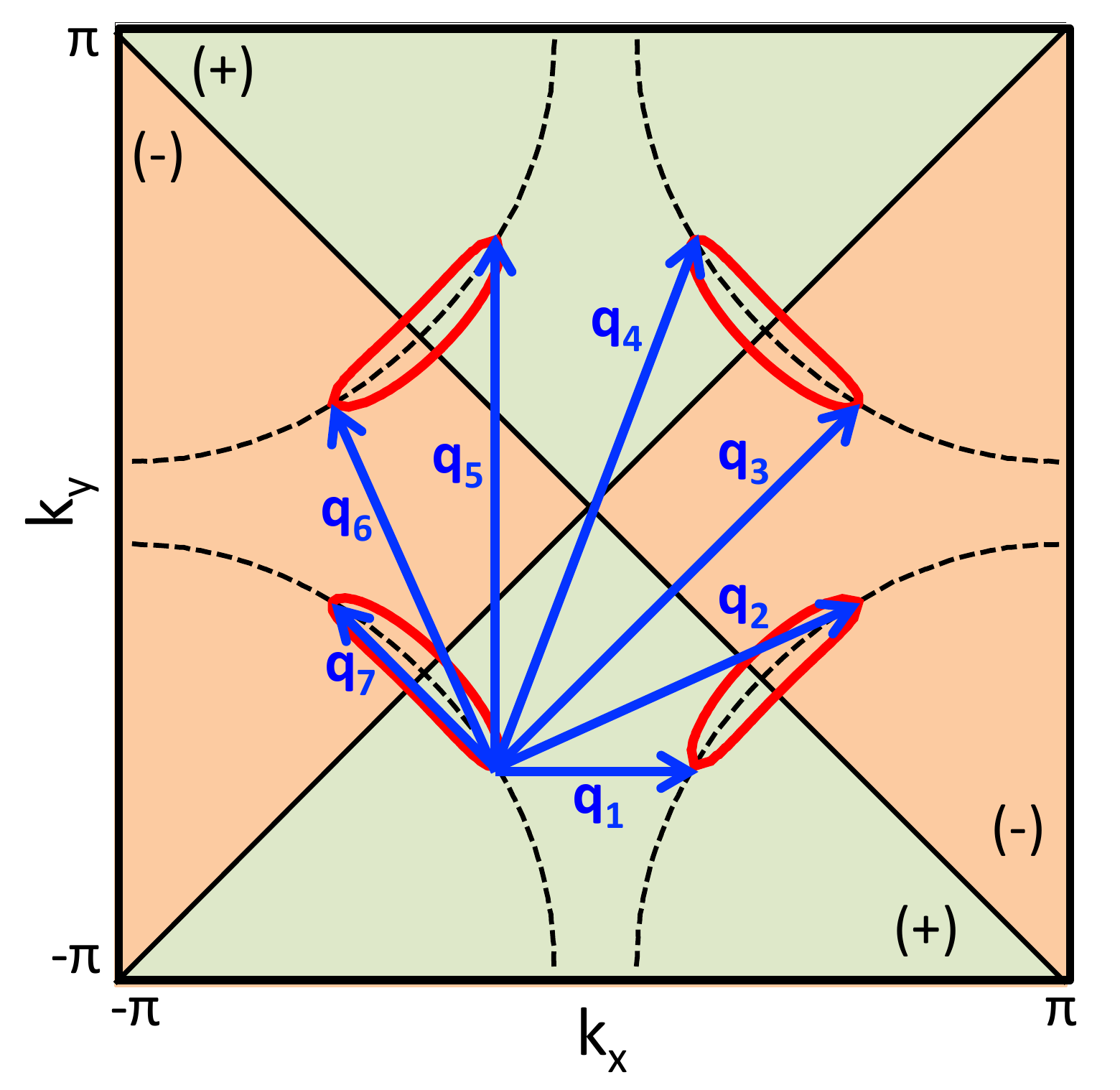}
\caption{\label{fig:octet} {\bf The octet model.} Scattering is dominated by wavevectors ${\bf q}_1$-${\bf q}_7$ which connect the eight points at the ends of banana-shaped contours of constat energy (red curves) centered on the normal state Fermi surface (dashed line). The phase of the $d$-wave superconducting order parameter is shown in green (positive) and orange (negative).}
\end{figure}

The superconducting QPI pattern in the cuprates is dominated by scattering wavevectors ${\bf q}_i$ predicted by the octet model.~\cite{wang_and_lee,dwavesea} The octet model is derived by noting that for a $d$-wave superconductor, the quasiparticle dispersion exhibits  banana-shaped contours of constant energy (CCE). The momentum space LDOS is maximal at the tips of the CCE, where the curvature is strongest.  Quasiparticle scattering is  dominated by wavevectors ${\bf q}_1-{\bf q}_7$ that connect the endpoints of the CCE, depicted in Fig.~\ref{fig:octet}. The presence of particle-hole symmetric wavevectors dispersing according to the octet model is evidence that an observed gap is due to superconductivity; the shape of the underlying Fermi surface and the momentum dependence of the  gap can be determined from this dispersion. 
 
 The evolution of the QPI intensity with bias energy has recently been a subject of discussion in the literature. FT-STS experiments have observed that the  intensity at some octet ${\bf q}$-vectors disappears when the tips of the banana-shaped CCE cross the antiferromagnetic zone boundary (AFZB), which has been interpreted as a signature of the loss of QPI, or more generally, the loss of quasiparticles.~\cite{Kohsaka_How_CP_vanish} In contrast, recent ARPES measurements observed well defined quasiparticles all the way from the node to the antinode, and it has been argued that the extinction of QPI at the AFZB observed in FT-STS can be attributed simply to the momentum dependence of  impurity scattering.~\cite{Inna} Additionally, the presence of a competing spin DW order coexisting with superconductivity can explain the disappearance of QPI beyond the AFZB.~\cite{Andersen}
 
 In addition to dispersing octet ${\bf q}$-vectors from superconducting QPI, static quasi-periodic ``checkerboard" modulations have  been observed in both the superconducting and pseudogap phases.~\cite{hanaguri_cb, hoffman_cb, McElroy_qpi2, Vershinin, Wise, Wise_cdw, Kohsaka_glass,McElroy_qpi1,Kohsaka_How_CP_vanish, davis_review, Parker}  Although the origin of these modulations remains unknown, there has been much discussion about their relation to a possible DW order in the pseudogap phase; scenarios that have been proposed include orbital current induced $d$-density waves,~\cite{chakravarty} one dimensional stripes,~\cite{kivelson} nematic order,~\cite{eun_ah_kim} short range charge order connected to nested parts of the Fermi surface in antinodal regions,~\cite{Wise_cdw, Li_checkerboard,CLi_checkerboard, Seibold_charge_order} and
 disorder-induced charge orders.~\cite{podolsky}

We present a detailed study of what can be learned from the Z-map in superconducting, DW, and coexisting phases.  In the first part of the paper we show that the evolution of the Z-map intensity with bias energy and doping reflects the underlying particle-hole asymmetry of the bandstructure.  We also compare the QPI patterns from different impurity types, and show how the Z-map can distinguish between them.  In the second half of the paper,  we consider DW order-- both long range and fluctuating-- and show that due to inherent particle-hole symmetry breaking, DW order produces a unique pattern in the Z-map.  As a result, the Z-map may be a useful in detecting and distinguishing between DW orders. 

The plan for the rest of this paper is the following: in Section II we discuss  general properties of the Z-map and its connection to particle-hole asymmetry. In Section III, using a single site impurity, we study the evolution of the QPI intensity with bias energy and doping, while in Section IV, we introduce an extended impurity with scattering treated self consistently. We investigate the signatures of DW order in the Z-map in Sections V-VII, considering the effect of long range and fluctuating DW orders, and their coexistence with superconductivity.   We conclude in Section VIII.

\section{Particle-hole asymmetry in the Z-map}
 Because the CCE in the superconducting state are particle-hole symmetric, the octet wavevectors are the same at positive and negative bias energy (${\bf q}_i(\omega)={\bf q}_i(-\omega)$), however, the QPI intensity (the magnitude of the LDOS modulations) at those wavevectors need not be.  Here, we consider the effect of  inherent particle-hole asymmetry in the underlying cuprate bandstructure  on the Z-map intensity; the difference between injecting and extracting electrons from a Mott insulator would further enhance this asymmetry in a real material.  The  Z-map intensity evolves with bias energy in part  because  $n({\bf r},\omega)$ is closer to particle-hole symmetric at low energies than at higher energies (where there is more asymmetry due to the proximity of the van Hove singularity).  Additionally, variation in the QPI intensity with doping can be understood by appealing to particle-hole asymmetry, as doping moves the van Hove singularity.  

In order to quantify these ideas, 
 we first decompose the Z-map into even and odd functions of bias energy, which makes clear the dependence  on the particle-hole asymmetry of the underlying bandstructure. The LDOS can be written as the sum of a uniform part and a part modulated by impurity scattering: $n({\bf r},\omega)=n_0(\omega)+\delta n({\bf r},\omega)$. When LDOS  modulations are weak, such that $|\delta n({\bf r},\omega)|<<|n_0(\omega)|$, the Fourier transform Z-map can be expanded as~\cite{Pereg_Barnea} 
\begin{eqnarray*}
\label{eq:zq}
Z({\bf q},\omega) &=& Z_0(\omega) \Big (\delta_{{\bf q},0}+\frac{\delta n({\bf q},\omega)}{n_0(\omega)}-\frac{\delta n({\bf q},-\omega)}{n_0(-\omega)} \Big ) \\
&=&Z_0(\omega) ( \delta_{{\bf q},0} + \delta Z({\bf q},\omega))
\end{eqnarray*}
where $Z_0(\omega)=n_0(\omega)/n_0(-\omega)$.
Decomposing the LDOS as  $\delta n({\bf q},\omega)=\delta n_{even}({\bf q},\omega)+\delta n_{odd}({\bf q},\omega)$, where $\delta n_{even}({\bf q},\omega)= \delta n_{even}({\bf q},-\omega)$ and $\delta n_{odd}({\bf q},\omega)= -\delta n_{odd}({\bf q},-\omega)$, the Z-map modulation can be expressed as
\begin{equation}
\delta Z({\bf q},\omega)=\alpha(\omega)\delta n_{odd}({\bf q},\omega)+\beta(\omega)\delta n_{even}({\bf q},\omega)
\label{eq:dZ}
\end{equation}
where
\begin{equation}
\alpha(\omega)=\frac{1}{n_0(\omega)}+\frac{1}{n_0(-\omega)} 
\label{eq:alpha}
\end{equation}
and
\begin{equation}
\beta(\omega)=\frac{1}{n_0(\omega)}-\frac{1}{n_0(-\omega)}. 
\label{eq:beta}
\end{equation}

The factors $\alpha(\omega)$ and $\beta(\omega)$ reflect the amount of particle-hole asymmetry in the underlying band structure.   For a particle-hole symmetric bandstructure,  $\alpha(\omega)=2/n_0(\omega)$ and $\beta(\omega)=0$, so the Z-map modulation depends only on $\delta n_{odd}({\bf q}, \omega)$:
\begin{equation}
\delta Z({\bf q},\omega)=\frac{2 \delta n_{odd}({\bf q},\omega)}{n_0(\omega)} 
\label{eq:zmap_phsymm}
\end{equation}
The cuprate bandstructure weakly breaks particle-hole symmetry for energies less than the gap maximum ($\omega<\Delta_0$), so in this case $\alpha(\omega)>>\beta(\omega)$.  As a result, the Z-map modulation is dominated by $\delta n_{odd}({\bf q}, \omega)$ in the cuprates. We use the expressions derived in this section frequently in our subsequent discussion.

\section{Single site impurity scattering}
Since QPI requires the presence of impurity scatterers, in this section we lay out the formalism for single impurity scattering, while in later sections we treat an extended impurity model.  We consider three impurity types in this paper: those that modulate the $d$-wave superconducting gap, the bond (nearest neighbor hopping) parameter, and the site energy.
The QPI pattern observed in FT-STS experiments arises from scattering from multiple impurity types, which can be difficult to disentangle. However, recent experimental~\cite{Hanaguri} and theoretical~\cite{Pereg_Barnea} work studying the effect of a magnetic field on QPI  has shown that vortices  induce gap inhomogeneities which can scatter quasiparticles. The magnetic field does not create additional scattering vectors, but enhances intensity at octet ${\bf q}$-vectors which originates from scattering from gap inhomogeneities, so these wavevectors now can be isolated and studied in detail.  

The theory of QPI from various impurity types has been studied by several authors. Calculations of QPI from single site and multiple impurities ~\cite{wang_and_lee, zhang_and_ting, zhang_and_ting2, Andersen,pb_calc} reproduce the locations of experimentally measured wavevectors, but the features in the QPI pattern are ``streak-like" rather than ``point-like" as in experiment.  Zhu $et. al$ ~\cite{zhu} studied QPI from multiple impurities, and highlighted the need for more realistic impurity models.  Other works have shown that many of the discrepancies between theory and experiment can be resolved by considering an extended long range impurity model,~\cite{nunner} or by including a spatially inhomogeneous chemical potential and superconducting gap.~\cite{dellanna}  However, the aim of this paper is to focus on the generic behavior of the Z-map, and we neglect more realistic impurity modeling.

We now review the formalism for single site impurity QPI calculations. While we neglect electronic correlations in this work, we do not expect them to change our overall conclusions, which largely are derived in a fairly straightforward way based on symmetry arguments.  In addition, we confine our considerations to the low temperature phases where there is strong experimental evidence that there are well-defined quasiparticles. The BCS Hamiltonian in the Nambu formalism is given by
\begin{equation}
\label{eq:h}
H=\sum_{\bf k} \psi_{\bf k}^\dagger \hat{\Lambda}_{\bf k} \psi_{\bf k},
\end{equation}
where $\psi_{\bf k}^\dagger=(c_{{\bf k} \uparrow}^\dagger, c_{-{\bf k}\downarrow}) $ and 
\[ \hat{\Lambda}_{\bf k}= \left ( \begin{array}{cc}
\epsilon_{\bf k} & \Delta_{\bf k} \\
\Delta_{\bf k}  & -\epsilon_{\bf k} \end{array} \right ). \]
Here $c_{{\bf k}\sigma}^\dagger$ creates an electron with momentum ${\bf k}$ and spin $\sigma$, $\epsilon_{\bf k}$ is the bandstructure, and for our considerations of the cuprates, we use the $d$-wave superconducting gap $\Delta_{\bf k}=\Delta_0(\cos k_x-\cos k_y)/2$. The Green's function is given by
\begin{eqnarray}
\label{eq:gf}
\hat{G}({\bf k},i\omega_n)&=&(i\omega_n\hat{I}-\hat{\Lambda}_{\bf k})^{-1} \\
&=& \frac{i\omega_n \hat{\tau}_0 + \Delta_{\bf k} \hat{\tau}_1 +\epsilon_{\bf k} \hat{\tau}_3}{(i\omega_n)^2-E_{\bf k}^2}, \nonumber
\end{eqnarray}
where $\hat{\tau}_0$ is the identity matrix, $\hat{\tau}_{1,3}$ are Pauli matrices, and $E_{\bf k}^2=\epsilon_{\bf k}^2+\Delta_{\bf k}^2$. 

On a lattice of $N$ sites, scattering from a single point-like impurity at site (0,0) is described by the Hamiltonian
\begin{equation}
H_{imp}=\sum_{{\bf r}}\psi_{\bf r}^\dagger \hat{T}_{{\bf r},{\bf 0}} \psi_{\bf 0}  + h.c.
\end{equation}
 where $\psi_{\bf r}^\dagger = (c_{{\bf r}\uparrow}^\dagger, c_{{\bf r}\downarrow})$ and  $\hat{T}_{{\bf r},{\bf 0}}=\hat{\tau}_{1,3}T_{{\bf r},{\bf 0}}$ is the scattering $T$-matrix. Contributions to the $T$-matrix from different impurity types are classified by how they modify electron parameters: bond modulations $\delta t$ and site-energy modulations $\delta \mu$ occur in the $\hat{\tau}_3$ channel, while $d$-wave superconducting gap modulations $\delta \Delta$ occurs in the $\hat{\tau}_1$ channel.

Superconducting gap and bond modulations modify the parameters on the four bonds immediately surrounding the impurity site.  On a lattice with spacing $a$ the real space $T$-matrix is given by
\begin{equation} 
 T_{{\bf r},{\bf 0}}=\delta m [\delta({\bf r}-a\hat{x}) \pm \delta({\bf r}-a\hat{y}) + \delta({\bf r}+a\hat{x}) \pm \delta({\bf r}+a\hat{y})]
\end{equation}
 where $\delta m=\delta t (\delta \Delta)$ is the amplitude of the modulation and the +(-) is for bond ($d$-wave gap) modulation. Site-energy modulation occurs only at the impurity site, for which $T_{{\bf r},{\bf 0}}=\delta \mu \delta({\bf r}-{\bf 0})$.  Transforming to momentum space, the $T$-matrix is
\begin{eqnarray}
\hat{T}^{g}_{{\bf k},{\bf k}+{\bf q}} &=&  \delta \Delta \hat{\tau}_1 (\Delta_{\bf k}+ \Delta_{{\bf k}+{\bf q}}) \\
\hat{T}^{b}_{{\bf k},{\bf k}+{\bf q}}& =& \delta t \hat{\tau}_3 ( t_{\bf k}+ t_{{\bf k}+{\bf q}}) \\
\hat{T}^{s}_{{\bf k},{\bf k}+{\bf q}}& =& \delta \mu \hat{\tau}_3 
\end{eqnarray}
for gap, bond, and site energy modulations, respectively. The definition of $\delta \Delta_{\bf k}$ was given above,  and $ t_{\bf k} =  \cos k_x+\cos k_y$. 

From the momentum space form of the T-matrix, we can already make two important observations about the energy dependence of the QPI intensity.~\cite{Inna}  First, $\hat{T}^{b}_{{\bf k},{\bf k}+{\bf q}}=0$ for  all octet ${\bf q}$-vectors  when ${\bf k}$ and ${\bf k}+{\bf q}$ lie along the AFZB. Therefore, for impurities that modulate the bond parameter, extinction of QPI at the AFZB is simply due to the momentum-space form of  the $T$-matrix, rather than a loss of quasiparticles.  Including next-nearest-neighbor bond modulations lifts the complete extinction of QPI intensity at the AFZB, although these modulations have a smaller amplitude.  Additionally, for superconducting gap modulations, $\hat{T}^{g}_{{\bf k},{\bf k}+{\bf q}} =0$  for octet ${\bf q}$-vectors that connect points of opposite gap phase (${\bf q}_2, {\bf q}_3, {\bf q}_6, {\bf q}_7)$.  All wavevectors are present for site energy modulations.

\subsection{Born scattering in a particle-hole symmetric band}

We now classify the different impurity types by whether they create an LDOS modulation that is even or odd with respect to bias energy, for the simple case of scattering in a particle-hole symmetric band.  This simple classification scheme will help us understand the behavior of $\delta Z({\bf q},\omega)$ in the cuprates, where this classification still holds, but only approximately.  The symmetry of the LDOS modulation due to superconducting gap and site energy variations already was described in  Ref. \onlinecite{shoucheng} using the Bogoliubov-de Gennes equations for a $d$-wave superconductor with classical phase fluctuations. We take a different approach: we derive the LDOS modulation due to scattering from a single  impurity site using a momentum space $T$-matrix formalism.

 The LDOS modulation for scattering by wavevector ${\bf q}$ is~\cite{dwavesea}
\begin{eqnarray}
\label{eqn:dn}
\delta n({\bf q},\omega)=&& \int \frac{d^2 {\bf k}}{(2 \pi)^2} \\
&& Im[\hat{G}({\bf k},\omega+i\delta)\hat{T}_{{\bf k},{\bf k}+{\bf q}}\hat{G}({\bf k}+{\bf q},\omega+i\delta)]_{11}\nonumber
\end{eqnarray}
where $\hat{G}({\bf k},\omega+i\delta)$ is the analytic continuation of the Nambu Green's function to real frequencies. This expression can be separated into parts that are even and odd with respect to bias energy $\omega$:
\begin{eqnarray}
\delta n({\bf q},\omega)&=&  \int  \frac{d^2 {\bf k}}{(2 \pi)^2}\frac{\delta(\omega-E_{\bf k})+\delta(\omega+E_k)}{E_{{\bf k}+{\bf q}}^2-\omega^2} \times \nonumber\\
&&\delta m[f_1({\bf k},{\bf q}) + f_2({\bf k},{\bf q})/\omega]
\end{eqnarray}
where $\delta m$, $f_1({\bf k},{\bf q})$ and  $f_2({\bf k},{\bf q})$ depend on the type of scatterer.  For gap modulation
\begin{eqnarray}
f_1^g({\bf k},{\bf q})&=&(\Delta_{\bf k}+\Delta_{{\bf k}+{\bf q}})^2 \\ \nonumber
f_2^g({\bf k},{\bf q})&=& (\Delta_{\bf k}+\Delta_{{\bf k}+{\bf q}})(\Delta_{\bf k}\epsilon_{{\bf k}+{\bf q}} + \Delta_{{\bf k}+{\bf q}}\epsilon_{\bf k}),
\end{eqnarray}
while for bond modulation
\begin{eqnarray}
f_1^b({\bf k},{\bf q})&=&(t_{\bf k}+t_{{\bf k}+{\bf q}})(\epsilon_{\bf k}+\epsilon_{{\bf k}+{\bf q}}) \\ \nonumber
f_2^b({\bf k},{\bf q})&=& (t_{\bf k}+t_{{\bf k}+{\bf q}})(\omega^2+\epsilon_{\bf k}\epsilon_{{\bf k}+{\bf q}}-\Delta_{\bf k}\Delta_{{\bf k}+{\bf q}}),
\end{eqnarray}
and for site energy modulation
\begin{eqnarray}
f_1^s({\bf k},{\bf q})&=&\epsilon_{\bf k}+\epsilon_{{\bf k}+{\bf q}}\\ \nonumber
f_2^s({\bf k},{\bf q})&=&\omega^2+\epsilon_{\bf k}\epsilon_{{\bf k}+{\bf q}}-\Delta_{\bf k}\Delta_{{\bf k}+{\bf q}}.
\end{eqnarray}
Noting that for a particle-hole symmetric bandstructure, $\epsilon_{{\bf k}+{\bf Q}}=-\epsilon_{\bf k}$ for ${\bf Q}=(\pi,\pi)$, it can be shown that
\begin{eqnarray}
f_2^g({{\bf k},{\bf q}})&=&0\\ \nonumber
f_2^b({{\bf k},{\bf q}})&=&0\\ \nonumber
f_1^s({{\bf k},{\bf q}})&=&0
\end{eqnarray}
so that for superconducting gap and bond modulations, the LDOS modulation is purely even with respect to bias energy, while for site energy modulation, the LDOS modulation is purely odd. 

As a result, since $\delta Z({\bf q},\omega)$ depends only on $\delta n_{odd}({\bf q},\omega)$  for a particle-hole symmetric bandstructure, only site energy modulations contribute to the spatially varying Z-map.
When p-h symmetry is broken, all three types of impurities cause both odd and even modulations in the LDOS. 

\subsection{Born scattering in a particle-hole asymmetric band}

Having classified the LDOS modulations in a particle-hole symmetric band, we now turn to a particle-hole asymmetric bandstructure describing the cuprates:
\begin{eqnarray}
\epsilon_{\bf k}&=&-2t(\cos k_x+\cos k_y)-4t'\cos k_x \cos k_y \nonumber\\
&&-2t''(\cos 2 k_x + \cos 2 k_y)- \nonumber \\
&&4t'''(\cos 2 k_x \cos k_y+\cos k_x \cos 2 k_y)-\mu
\label{eq:bandstr}
\end{eqnarray}
 where $t$, $t'$, $t''$, $t'''$, $ = $ 0.22, -0.034315, 0.035977, -0.0071637 eV respectively and the chemical potential $\mu$ is adjusted to control the filling. This bandstructure is obtained from a tight binding fit to ARPES data on nearly optimally doped Pb-Bi2201 described in Ref.~\onlinecite{ruihua}.  The $d$-wave superconducting gap maximum is set to  $\Delta_0$=35 meV.
The uniform DOS $n_0(\omega)$ for this band structure in the superconducting state is shown in Figure~\ref{fig:dos}(a).  
 
Fig.~\ref{fig:dos}(b) shows the factors $\alpha(\omega)$ and $\beta(\omega)$ (defined in Eqs.~\ref{eq:alpha} and~\ref{eq:beta}), which control the contributions of $\delta n_{odd}({\bf q},\omega)$ and $\delta n_{even}({\bf q},\omega)$ to $\delta Z({\bf q},\omega)$, respectively.  The term $\alpha(\omega)$ is an order of magnitude larger than $\beta(\omega)$, strongly enhancing the contribution from $\delta n_{odd}$.  Since $n_0(\omega)$ increases with energy $|\omega|$, especially at negative bias due to proximity to the van Hove singularity, $\alpha(\omega)$ decreases.  Thus the Z-map intensity should decrease with increasing energy, unless the strength of $\delta n_{odd}({\bf q},\omega)$ increases markedly.

\begin{figure}
\includegraphics[width=0.48\textwidth]{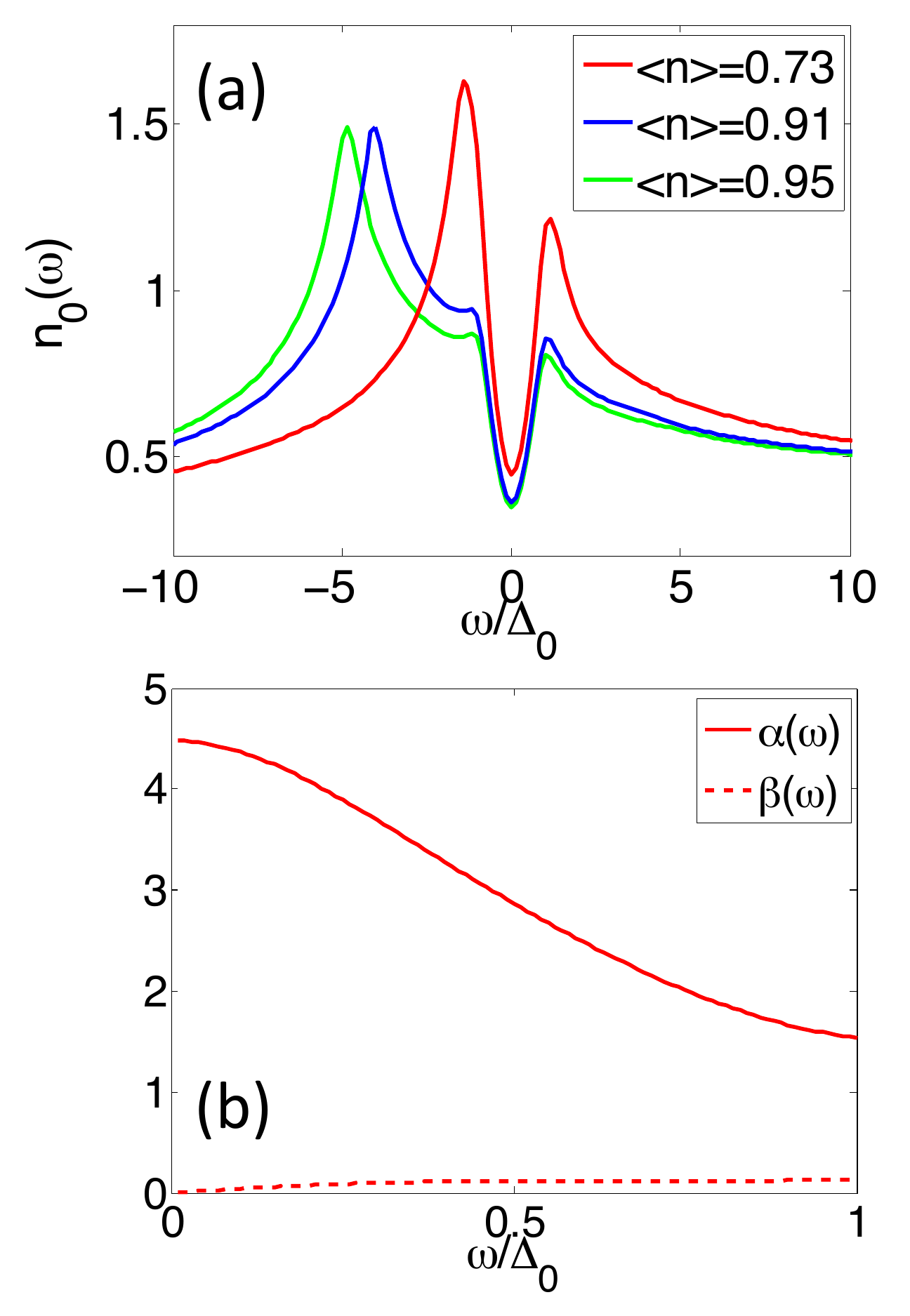}
\caption{\label{fig:dos} (a) {\bf Density of states} for the band structure defined in Eq.~\ref{eq:bandstr}.   (b) $\alpha(\omega)$ and $\beta(\omega)$ defined in Eqs.~\ref{eq:alpha} and~\ref{eq:beta}. for $<n>$=0.73.}
\label{fig:dos}
\end{figure}

\begin{figure}
\includegraphics[width=0.48\textwidth]{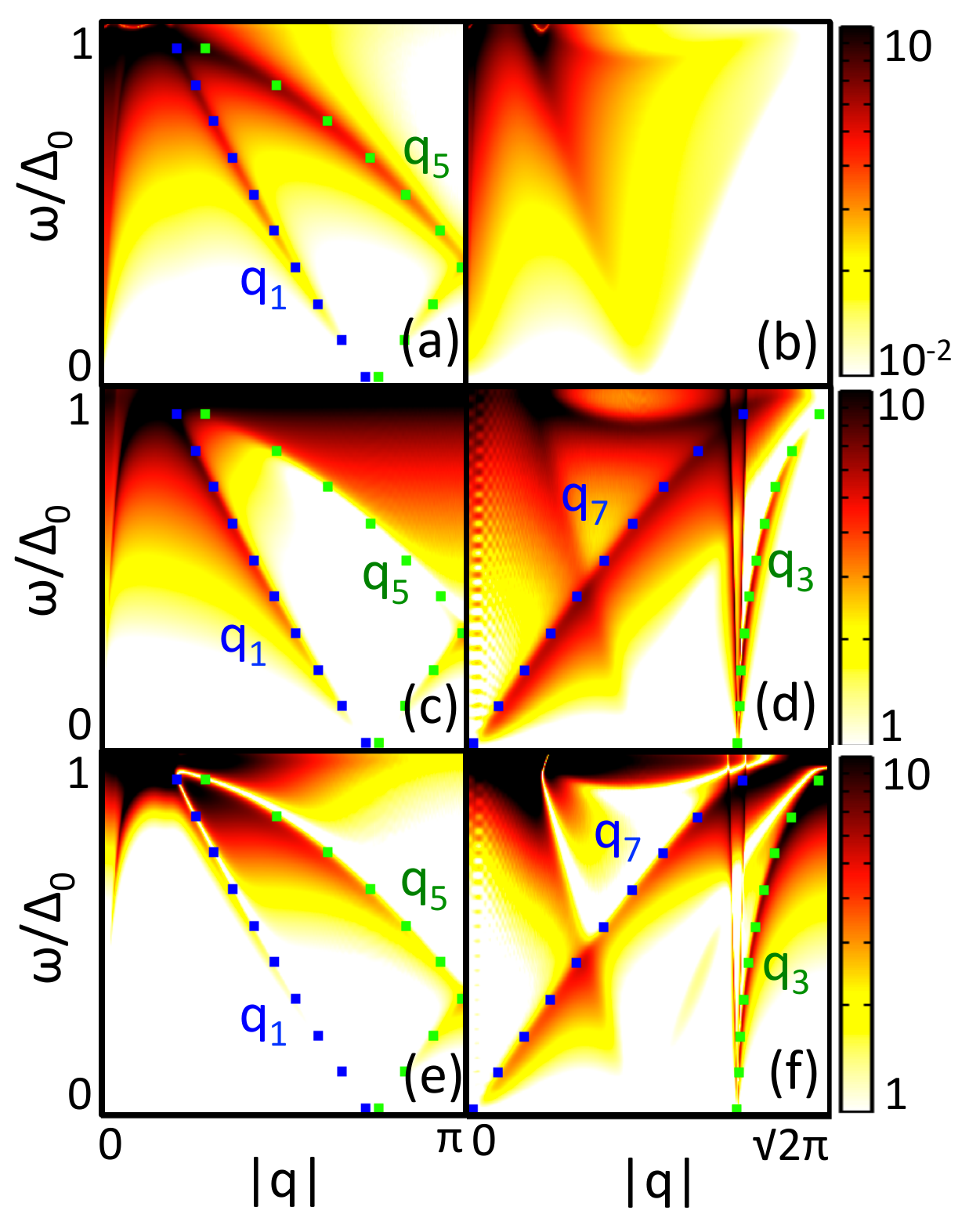}
\caption{\label{fig:dn_q_w} {\bf Odd part of the LDOS modulation} $|\delta n_{odd}({\bf q},\omega)|$ as a function of scattering wavevector ${\bf q}$ and energy $\omega$ for (a, b) superconducting gap modulation, (c, d)  bond modulation, and (e, f) site energy modulation. The wavevector ${\bf q}$ for the left column (panels a, c, and e) is along the anti-nodal cut (${\bf q}=(0,0)\rightarrow(\pi,0)$), while in the right column (panels b, d, f) it lies along the nodal cut (${\bf q}=(0,0)\rightarrow(\pi,\pi)$). The boxes mark the location of the ${\bf q}$-vectors as calculated in the octet model that lie along the antinodal (${\bf q}_1$ and ${\bf q}_5$) or nodal cuts (${\bf q}_3$ and ${\bf q}_7$).  Intensities are in arbitrary units.}
\end{figure}

 Figure~\ref{fig:dn_q_w} shows $|\delta n_{odd}({\bf q},\omega)|$, as a function of scattering wavevector ${\bf q}$ and energy $\omega$ for superconducting gap  (panels a-b), bond  (c-d), and site energy  (e-f) modulations. Two Brillouin zone cuts are shown, which highlight different octet ${\bf q}$-vectors (refer to Fig.~\ref{fig:octet} for the definitions of these vectors): ${\bf q}_1$ and ${\bf q}_5$ in an anti-nodal $(0,0)\rightarrow (\pi,0)$ cut  (panels a, c, and e), and  ${\bf q}_3$ and ${\bf q}_7$  along a nodal $(0,0)\rightarrow (\pi,\pi)$  cut  (panels b, d, and f).

In the anti-nodal cut, ${\bf q}_1$ and ${\bf q}_5$ start out with negligible intensity at low energy, and then increase in intensity as energy increases for all three types of modulation (panels a, c, and e).  The effect is most pronounced for superconducting gap modulation, where the intensity increases by three orders of magnitude. 
For the nodal cut,  wavevectors ${\bf q}_3$ and ${\bf q}_7$ are not present for superconducting gap modulation (panel b), due to the momentum dependence discussed at the beginning of this section.   For bond modulation (panel d), ${\bf q}_3$ and ${\bf q}_7$ both start out with low intensity at small energies, and then quickly grow in intensity. This intensity eventually starts to decrease with increasing energy.   The behavior of ${\bf q}_3$ and ${\bf q}_7$ in the case of site energy modulation (panel f) is qualitatively similar, although both wavevectors continue to gain intensity at high energies.

In summary, from simple DOS arguments, the Z-map intensity $\delta Z({\bf q},\omega)$ at all  octet wavevectors should decrease with increasing $|\omega|$ due to the reduction in $\alpha(\omega)$.  However, a strong increase in $\delta n_{odd}({\bf q},\omega)$, for example, at large $|\omega|$ in the case of  gap modulation, or a peak at intermediate  $|\omega|$ in the case of bond and site energy modulations, could outweigh this behavior, depending on the overall strength of the impurity scattering.  The overall evolution of the Z-map intensity will depend on the interplay of these two effects, which we will illustrate in the next section.

\section{Scattering from an impurity patch in self-consistent $T$-matrix approach}
To more accurately describe impurity scattering and better compare to FT-STS data, we consider the effect of an extended impurity.~\cite{Inna,Fang}  We use an extended patch, which is embedded in a finite periodic lattice, and calculate the $T$-matrix self-consistently in real space. The real space Green's function is given by
\begin{equation}
\hat{G}({\bf r},i\omega_n)=\frac{1}{N}\sum_{\bf k} \hat{G}({\bf k},i\omega_n) e^{i {\bf k} \cdot {\bf r}}
\end{equation}
where $\hat{G}({\bf k},i\omega_n)$ is the Nambu Green's function given in Eq.~\ref{eq:gf}.  The initial real space $T$-matrix $\hat{T}_0({\bf r_1},{\bf r_2})$ gives the amplitude for scattering between two adjacent sites in the patch for  superconducting gap or bond modulation, or the amplitude for on-site scattering for site energy modulation. The $T$-matrix for scattering between any points ${\bf r}_1$ and ${\bf r}_2$ in the impurity patch  is then determined self-consistently from 
\begin{equation}
\hat{T}({\bf r}_1, {\bf r}_2)=\hat{T}_0({\bf r}_1, {\bf r}_2) +\sum_{{\bf r},{\bf r}'} \hat{T}_0({\bf r}_1,{\bf r}')\hat{G}({\bf r}'-{\bf r})\hat{T}({\bf r},{\bf r}_2)
\end{equation}
(the $\omega$ dependence of the Green's function has been suppressed). The LDOS modulation in real space due to quasiparticle scattering is given by
\begin{equation}
\delta n({\bf r},\omega)=-\frac{1}{\pi} Im \sum_{{\bf r}_1,{\bf r}_2} \Big ( \hat{G}({\bf r}-{\bf r}_1)\hat{T}({\bf r}_1,{\bf r}_2)\hat{G}({\bf r}_2-{\bf r}) \Big )_{11}
\end{equation}
and the real space Z-map is
\begin{equation}
 Z({\bf r},\omega)=\frac{n_0(\omega)+\delta n({\bf r},\omega)}{n_0(-\omega)+\delta n({\bf r},-\omega)}.
\end{equation}
The Fourier transform (FT) Z-map is then
\begin{equation}
Z({\bf q},\omega)=\frac{1}{N}\sum_{\bf r} e^{-i{\bf q}\cdot{\bf r}}Z({\bf r},\omega).
\label{eq:zmap_q}
\end{equation}

\begin{figure*}
\begin{center}
\includegraphics[width=\textwidth]{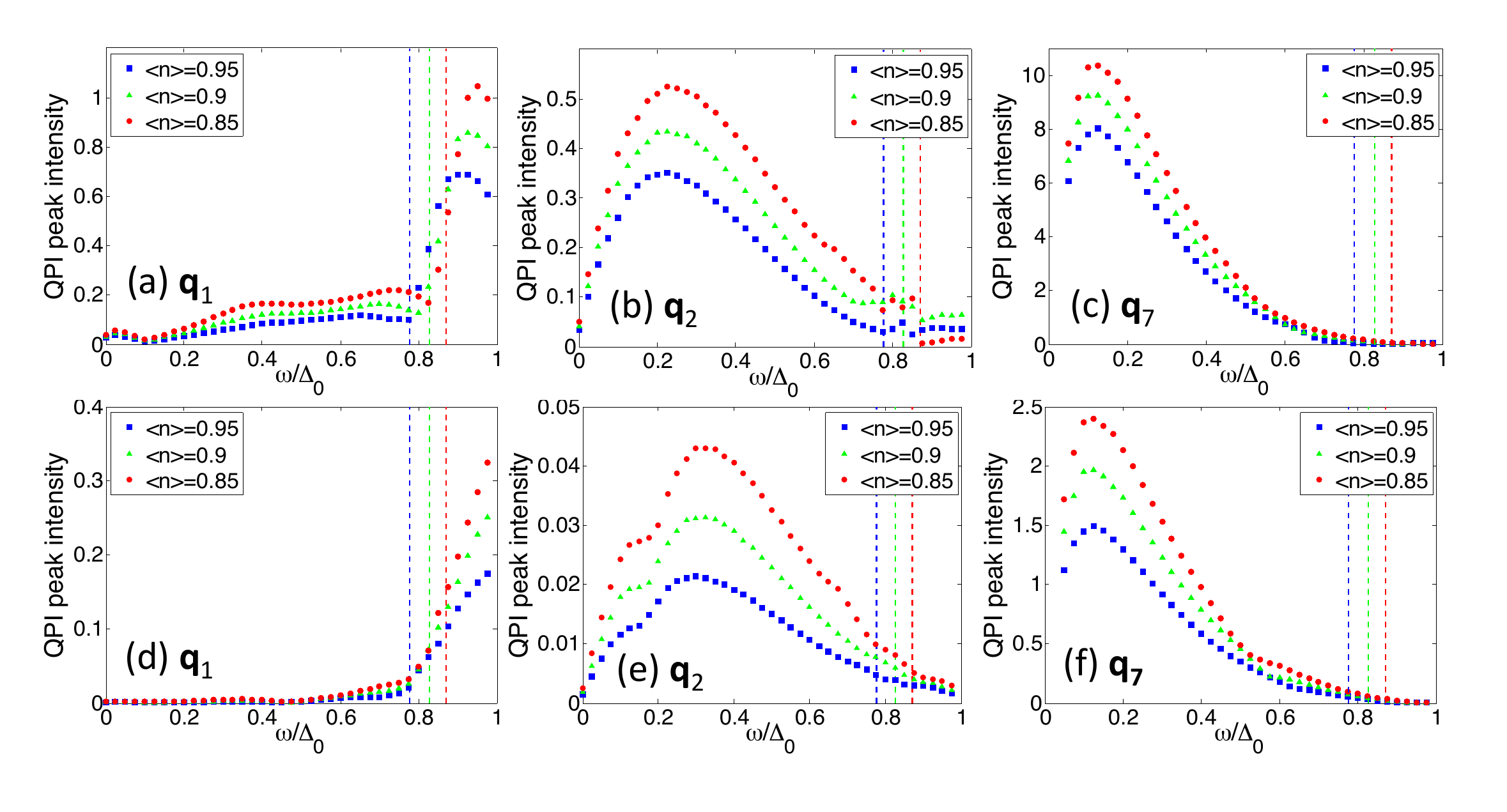}
\end{center}
\caption{\label{fig:qpi} {\bf FT Z-map intensity} for three different levels of hole doping. The QPI intensity in panels (a)-(c) is extracted from a Z-map calculation with a ``patch" impurity that modulates the bond and superconducting gap parameters. The energy at which the ends of the CCE cross the AFZB for each doping level are denoted by dashed lines. The QPI intensity in panels (d)-(f) is extracted from a calculation with a ``patch" impurity that modulates the site energy. The behavior is qualitatively similar to that of bond/gap modulation. The QPI intensity is in arbitrary units.}
\end{figure*}

The figures shown in the rest of this paper were made using a $9\times 9$ extended impurity embedded in the center of a $512\times 512$ real space lattice. The magnitude of the modulation parameters in the initial $T$-matrix are a maximum on the bonds in the center of the patch, and decrease in magnitude within a ``Gaussian-like" envelope, in order to minimize the effect of ``ringing" that would occur from an abrupt fall-off of the $T$-matrix at the patch boundaries. Since the impurity in our calculation is embedded in the center of the real space lattice, it creates an LDOS modulation $\delta n({\bf r},\omega)$, and hence a real space Z-map $Z({\bf r},\omega)$ that is $C4$ symmetric. As can be seen from Eq.~\ref{eq:zmap_q}, the Fourier transform of a $C4$ symmetric quantity (here $Z({\bf q},\omega)$) is purely real. 

Fig.~\ref{fig:qpi} shows the FT Z-map $Z({\bf q},\omega)$  for a representative set of octet ${\bf q}$-vectors at three levels of hole doping, for an impurity that modulates the superconducting gap and bond parameters (panels a-c), or site energy (panels d-f). The value of $\mu$  in Eq.~\ref{eq:bandstr} is varied to control the hole doping, and the superconducting gap maximum is set to $\Delta_0=35$ meV in all cases. Changing the doping in the system moves the van Hove singularity closer to the superconducting gap edge (more hole doped) or further away (less hole doped), as shown in Fig.~\ref{fig:dos}(a), which affects the  particle-hole asymmetry. As a result, for all octet ${\bf q}$-vectors, the FT Z-map intensity becomes larger as the system is more strongly hole-doped. 

The FT Z-map intensity at  ${\bf q}_1$ (Fig.~\ref{fig:qpi}a), ${\bf q}_4$,  and ${\bf q}_5$ (not shown, but qualitatively similar) has low intensity until near the energy at which the ends of the CCE cross the AFZB, when the intensity rapidly increases. Recall that the QPI intensity is extinguished at the AFZB for bond modulating impurities, so this increase is solely due to the LDOS modulation from superconducting gap variation (see Fig.~\ref{fig:dn_q_w}).   The FT Z-map intensity at  ${\bf q}_2$,  ${\bf q}_7$ (Fig.~\ref{fig:qpi} b-c), ${\bf q}_3$,  and ${\bf q}_6$ (not shown, but qualitatively similar), peaks and then decreases with energy due to the decrease in $\alpha(\omega)$, until the intensity is completely extinguished at the AFZB for symmetry reasons as discussed in Section III.  

The behavior of the octet ${\bf q}$-vectors in the case of site energy modulation (Fig.~\ref{fig:qpi} d-f), is qualitatively similar to that for bond/gap modulation.  There is no symmetry reason for the extinction of FT Z-map intensity at the AFZB, so ${\bf q}_2$ evolves smoothly through this boundary.  Because the behavior of the Z-map intensity for site energy modulation and bond/gap modulation are similar, the rest of the FT Z-map figures shown in the paper are computed with only bond/gap modulations. A comparison of the FT Z-maps for these three types of impurity scatterers, all considered separately, has also appeared in Ref.~\onlinecite{Inna}.

To summarize our results on the energy dependence of the FT Z-map intensity, we have shown that the energy and doping evolution of the octet ${\bf q}$-vectors reflect the particle-hole asymmetry in the underlying bandstructure.  We also have shown that the different types of impurity scatterers give different ${\bf q}$-dependent QPI intensities, so that the intensity of QPI at certain wavevectors could be used to identify the type of underlying impurity scattering. Namely, ${\bf q}$-vectors that rise quickly from negligible intensity at the AFZB can be associated with gap modulating impurities, while those that peak and then extinguish at the AFZB are from bond modulations; site energy modulating impurities can exhibit both of these behaviors.  While these behaviors offer a way to distinguish between impurity types, disentangling the effects of multiple impurity types in one system is still difficult.  In this case, techniques that highlight certain scattering types, such as the use of magnetic field to enhance octet ${\bf q}$-vectors arising from gap inhomogeneity scattering, as demonstrated in Refs.~\onlinecite{Pereg_Barnea} and ~\onlinecite{Hanaguri}, are especially relevant.

\begin{figure}
\includegraphics[width=0.48\textwidth]{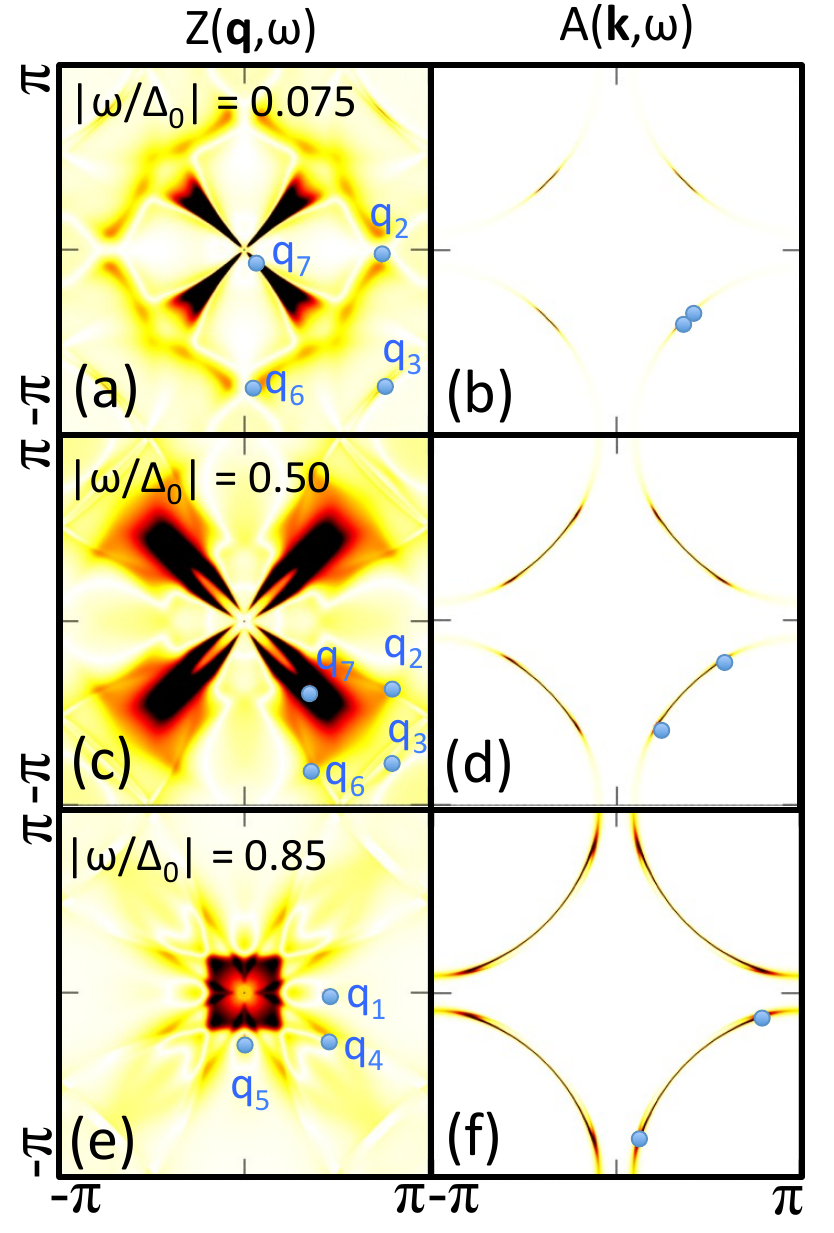}
\caption{\label{fig:zmap-sc} {\bf $d$-wave superconductivity.} $Z({\bf q},\omega)$ (left column) and spectral function  $A({\bf k},\omega)$ (right column) for $|\omega/\Delta_{0}|=0.075$ (a and b), $|\omega/\Delta_{0}=0.5|$ (c and d), and $|\omega/\Delta_{DW}=0.85|$ (e and f).  Here the magnitude of the $d$-wave superconducting gap is $\Delta_{0}=35$ meV.  Blue circles mark the locations of the octet  ${\bf q}$-vectors  on $Z({\bf q},\omega)$, and the ends of the CCE on $A({\bf k},\omega)$, respectively.   }
\end{figure}

In order to facilitate the comparison of FT Z-maps with DWs in the subsequent sections to the case of pure superconductivity discussed here, we plot the FT Z-map $Z({\bf q},\omega)$ and spectral function $A({\bf k},\omega)$ at three bias energies in Figure~\ref{fig:zmap-sc}. In all subsequent FT Z-map plots in this paper (Figs.~\ref{fig:zmap-sc}, ~\ref{fig:lr_dw_zmap}, ~\ref{fig:dw_and_sc_zmap} --~\ref{fig:cb}), the filling is set to $<n>$=0.73 and the $Z({\bf q},\omega)$ intensity color scale for all panels in each figure is the same, although this scale has been adjusted between figures for clarity. Note that for each point in  Fig.~\ref{fig:qpi}, the QPI intensity at each octet ${\bf q}$-vector was extracted from FT Z-maps like these.  Peaks at wavevectors ${\bf q}_2$, ${\bf q}_3$, ${\bf q}_6$, and ${\bf q}_7$ are present at low energies and then extinguish as energy increases, while wavevectors ${\bf q}_1$, ${\bf q}_4$ and ${\bf q}_5$ appear only at high energy. The peaks all disperse according to the octet model.

\section{Long range density wave order}

\begin{figure}
\includegraphics[width=0.48\textwidth]{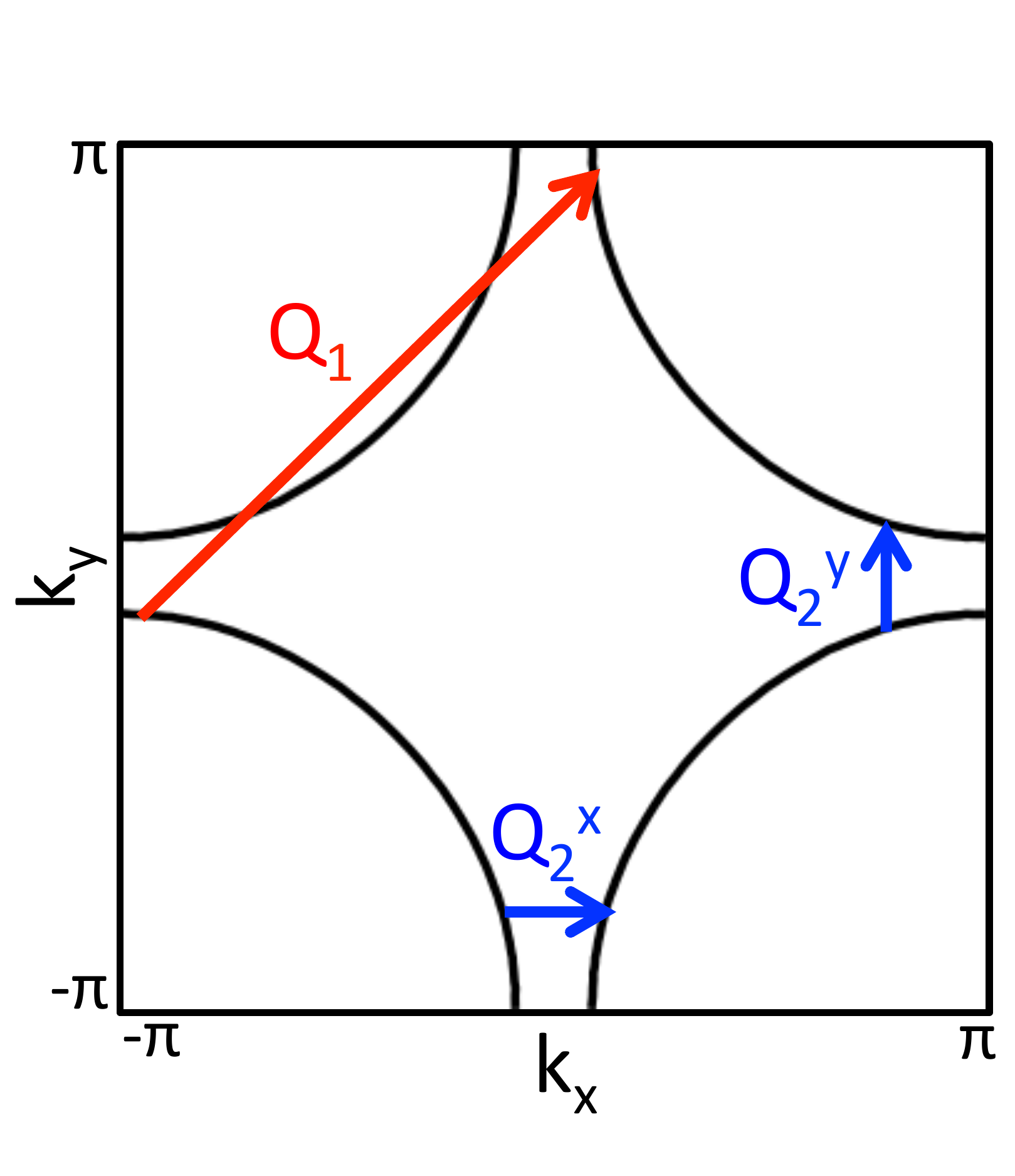}
\caption{\label{fig:dw_vec} {\bf DW ordering vectors}: commensurate ${\bf Q}_1 = (\pi,\pi)$ (red) and checkerboard ${\bf Q}_2^x=(0.26\pi, 0)$, ${\bf Q}_2^y=(0, 0.26\pi)$ (blue).}

\end{figure}

We now explore the signatures of DW order in the Z-map. This is motivated by the observation that the Z-map intensity is sensitive to particle-hole asymmetry, as discussed in the previous sections in the context of a particle-hole asymmetric bandstructure. Since DWs also break particle-hole symmetry, the Z-map may be a useful quantity to consider in DW systems.  The Z-map could be used to study any material system with a charge or spin order; for the case of the cuprates, it could be used to search for an underlying DW order in the pseudogap phase.  

Here we consider charge order, although spin orders could be treated in an analogous manner. We study two different DW orders in this work, illustrated in Fig.~\ref{fig:dw_vec}: commensurate ${\bf Q}_1=(\pi, \pi)$ order, and incommensurate checkerboard order.  The wavevectors of the checkerboard order, nesting the flat regions of the Fermi surface near the antinode for the bandstructure defined in Eq.~\ref{eq:bandstr}, are ${\bf Q}_2^x=(0.26\pi, 0)$, ${\bf Q}_2^y=(0, 0.26\pi)$.~\cite{Makoto} We focus on these two DW orders because the $(\pi, \pi)$ case is the simplest to treat, making the underlying physics most clear, while the checkerboard order has been proposed as a possible explanation for the checkerboard modulations in the cuprates.  The goal of the next few sections is to identify the signatures of these DW orders in the Z-map, and determine how to differentiate superconducting QPI from them.  In this section we focus on  long range DW order, while in subsequent sections we consider the coexistence of superconductivity with DWs, as well as fluctuating DW order.  

\begin{figure}

\includegraphics[width=0.48\textwidth]{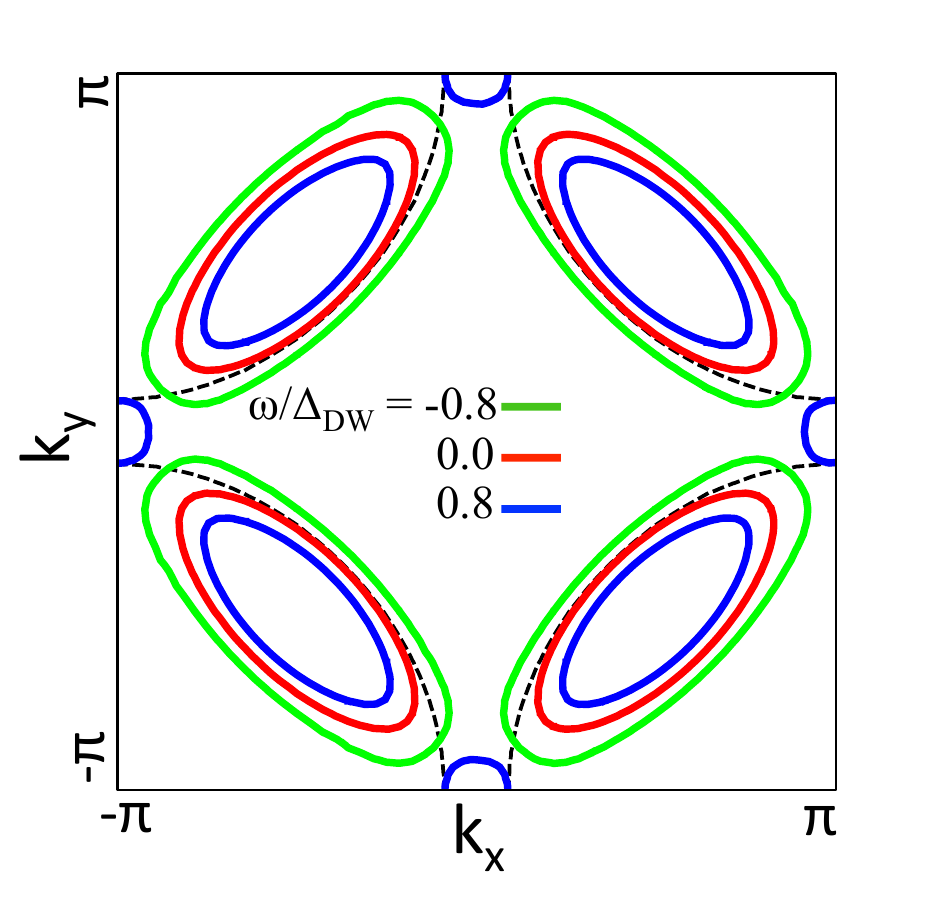}
\caption{\label{fig:dw_pipi} {\bf CCE for  ${\bf Q}=(\pi,\pi)$ long range DW order} with an isotropic gap $\Delta_{DW}$=100 meV. The normal state Fermi surface is shown with dashed lines.}
\end{figure}

We first consider a ${\bf Q}=(\pi,\pi)$ DW with an isotropic gap $\Delta_{DW}$.  The Hamiltonian is expressed in the Nambu formalism (Eq.~\ref{eq:h}), where now $\psi_{{\bf k}\sigma}=(c_{{\bf k}\sigma}, c_{{\bf k}+{\bf Q}\sigma})$ and 
\[ \hat{\Lambda}_{\bf k}= \left ( \begin{array}{cc}
\epsilon_{\bf k} & \Delta_{DW} \\
\Delta_{DW}  & \epsilon_{{\bf k}+{\bf Q}} \end{array} \right ), \]
and the sum in Eq.~\ref{eq:h} runs over the reduced Brillouin zone $|k_x|+|k_y| < \pi$. While the mean-field treatment of the DW order used in this paper certainly cannot fully explain the pseudogap in the underdoped cuprates, where electronic correlations, neglected in this work, are important, it can offer insight into generic signatures of DW order. The CCE are ellipses, as depicted in Fig.~\ref{fig:dw_pipi}. Because a DW order breaks particle-hole symmetry, unlike superconductivity, the CCE at bias energies $+\omega$ and $-\omega$ are of different sizes.  Since the CCE are not as sharply pointed as the superconducting banana-shaped CCE,  the maxima in $Z({\bf q},\omega)$ are more arc-like, as discussed below.  

We now turn to the signature of a ${\bf Q}=(\pi,\pi)$ DW in the FT Z-map, which is computed with the same self-consistent $T$-matrix formalism described in Section IV, except now the Green's function is derived from the Hamiltonian given in this section.  While in the superconducting state, maxima in $n({\bf r},\omega)$ correspond to minima in $n({\bf r},-\omega)$,  in the DW state the modulation patterns in $n({\bf r},\omega)$ and $n({\bf r},-\omega)$ are fundamentally different.  As a result, $Z({\bf r},\omega)$ does not enhance the strength of peaks as it does in the superconducting case, rather, separate peaks appear  corresponding to maxima (minima) in $n({\bf r},\pm\omega)$.  

\begin{figure}
\includegraphics[width=0.48\textwidth]{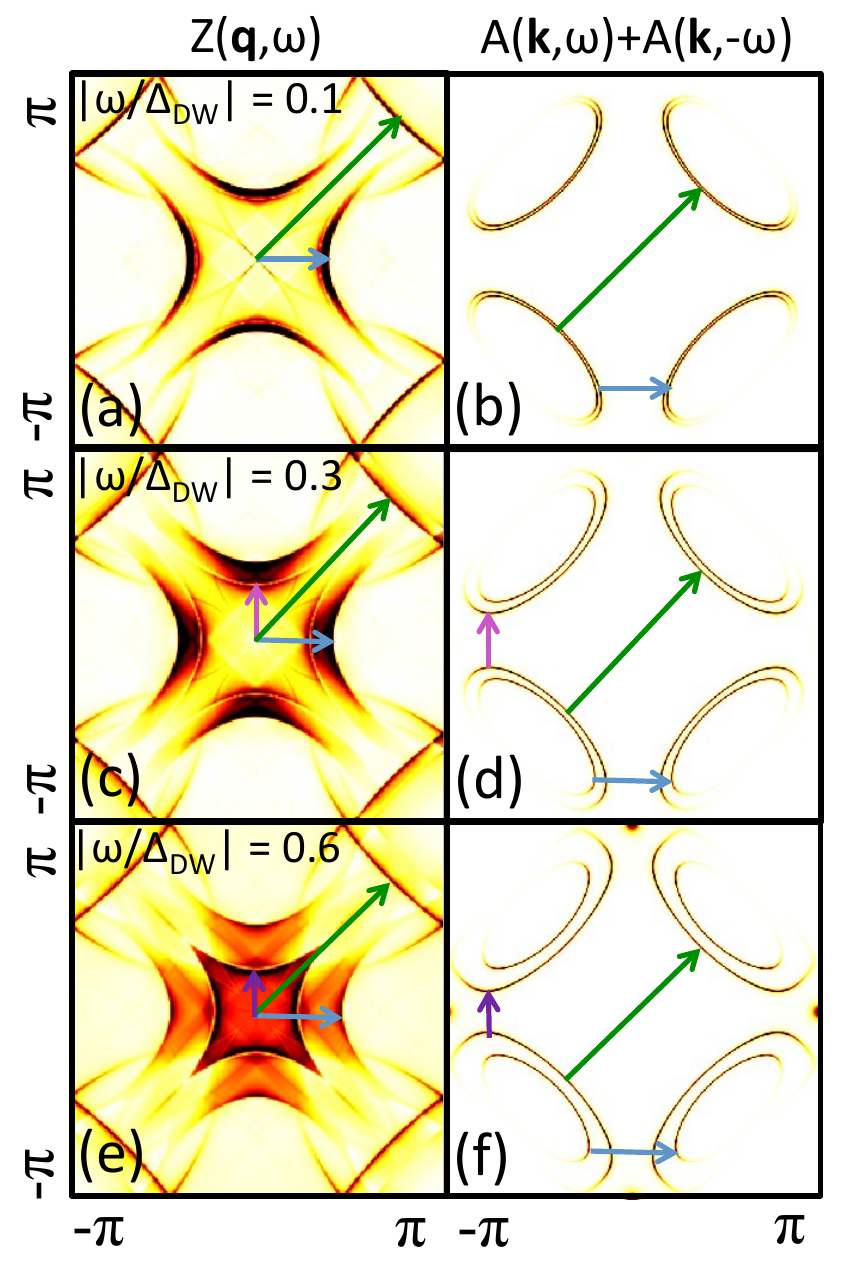}
\caption{\label{fig:lr_dw_zmap} {\bf Long range ${\bf Q}=(\pi,\pi)$ DW.} $Z({\bf q},\omega)$ (left column) and sum of spectral function at positive negative bias energies, $A({\bf k},\omega)+A({\bf k},-\omega)$ (right column) for $|\omega/\Delta_{DW}|=0.1$ (a and b), $|\omega/\Delta_{DW}|=0.3$ (c and d), and $|\omega/\Delta_{DW}|=0.6$ (e and f).  Here $\Delta_{DW}=100$ meV.  Peaks in $Z({\bf q},\omega)$ are marked with arrows color coded with the corresponding scattering vector in the spectral function.}
\end{figure}

The FT Z-map for a ${\bf Q}=(\pi,\pi)$ DW with gap $\Delta_{DW}=100$ meV  is shown in Fig.~\ref{fig:lr_dw_zmap} (panels a, c, and e) for three bias energies.  The sum of the spectral functions at positive and negative bias energy, $A({\bf k},\omega)+A({\bf k},-\omega)$, is displayed in panels b, d, and f.  By considering this sum, the difference between the spectral function at positive and negative bias energy is clear.  A ``double-arc" structure appears in $Z({\bf q},\omega)$ at all energies, and becomes more pronounced with increasing bias energy.  This structure arises because the CCE at positive and negative bias energy are of different sizes, and the size difference becomes more pronounced with increasing energy. The color coded arrows in Fig.~\ref{fig:lr_dw_zmap} mark areas of high intensity in $Z({\bf q},\omega)$, and the corresponding scattering vectors connecting regions of high intensity in the spectral function from which they arise.  At low energy (Fig.~\ref{fig:lr_dw_zmap}a, b), scattering is dominated by a wavevector that connects the ends of the ellipses and a wavevector that connects the ellipse centers.  Note that because the CCE at $\pm \omega$ are slightly different, there is a smearing of intensity in $Z({\bf q},\omega)$ (at the blue arrow).  The same two wavevectors dominate as energy increases; however, at intermediate energy (Fig.~\ref{fig:lr_dw_zmap}c, d), the single peak marked by the blue arrow in Fig.~\ref{fig:lr_dw_zmap}a now splits in two (blue and purple arrows), because the nesting vectors for the CCE at $\pm \omega$ are now significantly different, as shown in Fig.~\ref{fig:lr_dw_zmap}d.  This effect is enhanced with increasing energy (Fig.~\ref{fig:lr_dw_zmap}e, f). The appearance of these ``doubled" peaks in the Z-map is an indication of  DW order, rather than superconductivity.  We note that a doubling of the number of QPI octet wavevectors from particle-hole asymmetry has also been discussed in Ref.~\onlinecite{lee_stm}.  

For the case of checkerboard order, the Fermi surface is significantly reconstructed due to the many higher order harmonics associated with an incommensurate order.  Again, due to particle-hole symmetry breaking in the the DW phase, the CCE at $\pm \omega$ are different, so the ``doubling" of peaks in the FT Z-map, discussed for the $(\pi,\pi)$ DW, still holds. However, many peaks would appear in the FT Z-map due to the significant Fermi surface reconstruction, so it may be difficult to differentiate the ``doubled" peaks.

\section{Coexistence of superconductivity and long range density wave order}
  
We now consider superconductivity coexisting with long range ${\bf Q}=(\pi,\pi)$ DW order. Since the Z-map enhances the superconducting QPI peaks due to the coherence factors, while it creates a ``doubling" of features due to DW order, a system with both phases  will show an interplay of these two effects. Again, while here we focus on the particular case of the cuprates,  the Z-map is potentially useful for studying any material system with possible DW orders coexisting or competing with superconductivity.
      
\begin{figure}

\includegraphics[width=0.48\textwidth]{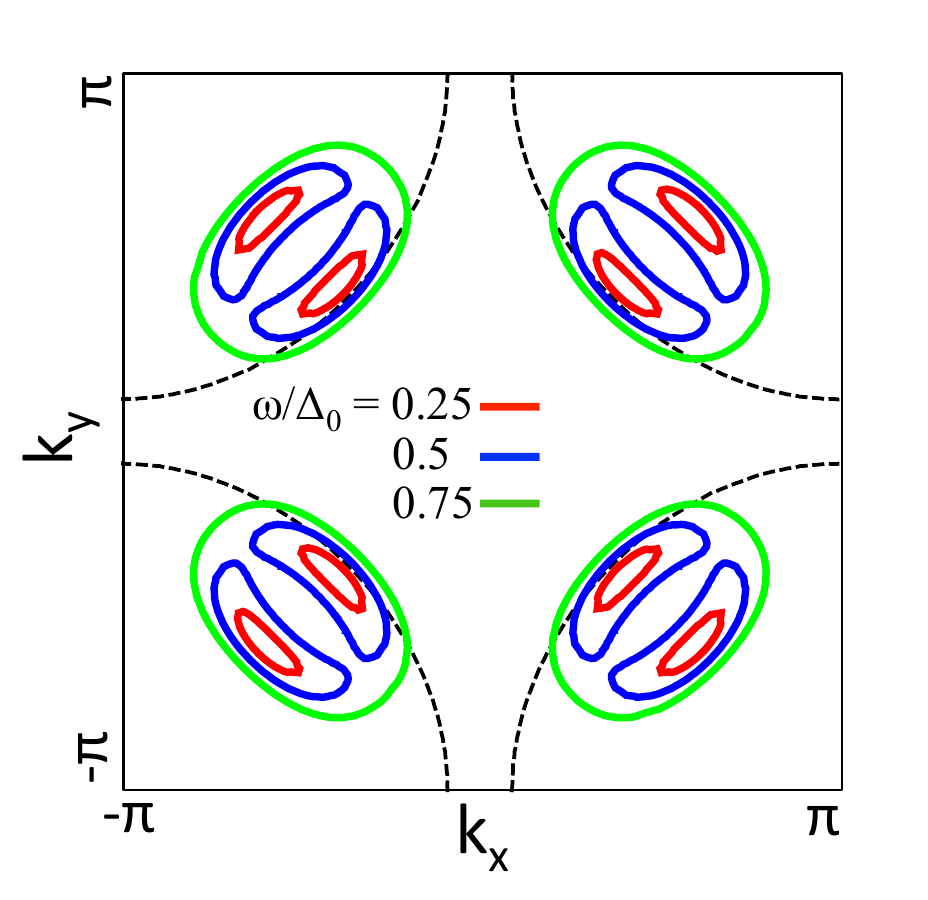}
\caption{\label{fig:dw} {\bf CCE for coexisting superconductivity and ${\bf Q}=(\pi,\pi)$ long range DW} with $\Delta_0=200$ meV and $\Delta_{DW}=250$ meV (values chosen for clarity). The normal state Fermi surface is shown with dashed lines.}
\end{figure}

The Nambu spinors in the Hamiltonian defined in Eq. \ref{eq:h} now have four components: $\psi_{\bf k}^\dagger=(c_{{\bf k} \uparrow}^\dagger,c_{{\bf k}+{\bf Q} \uparrow}^\dagger,c_{-{\bf k}\downarrow}, c_{-{\bf k}-{\bf Q}\downarrow}) $.  The matrix $\hat{\Lambda}_{\bf k}$ is $4\times4$:

\[ \hat{\Lambda}_{\bf k}= \left ( \begin{array}{cccc}
\epsilon_{\bf k} & \Delta_{DW} & \Delta_{\bf k} & 0 \\
\Delta_{DW}  & \epsilon_{{\bf k}+{\bf Q}} & 0 & -\Delta_{{\bf k}} \\
\Delta_{\bf k} & 0 & -\epsilon_{\bf k} & \Delta_{DW} \\
0 & -\Delta_{{\bf k}} & \Delta_{DW} & -\epsilon_{{\bf k}+{\bf Q}} \end{array} \right ) \]
where the sum in the Hamiltonian is over the reduced Brillouin zone $|k_x|+|k_y| < \pi$.  The CCE for coexisting superconductivity and $(\pi,\pi)$ DW order are shown in Fig.~\ref{fig:dw}. Due to the unit cell doubling in the DW phase, shadow CCE appear outside the reduced Brillouin zone, although these are suppressed in the spectral function due to the coherence factors.  Because ${\bf Q}=(\pi,\pi)$ does not nest the nodal region, at low energies, the CCE are banana-shaped as in the pure superconducting state.  Upon increasing energy, the banana-shaped CCE evolve into ellipses reminiscent of the CCE in the pure DW phase.

\begin{figure}
\begin{center}
\includegraphics[width=0.48\textwidth]{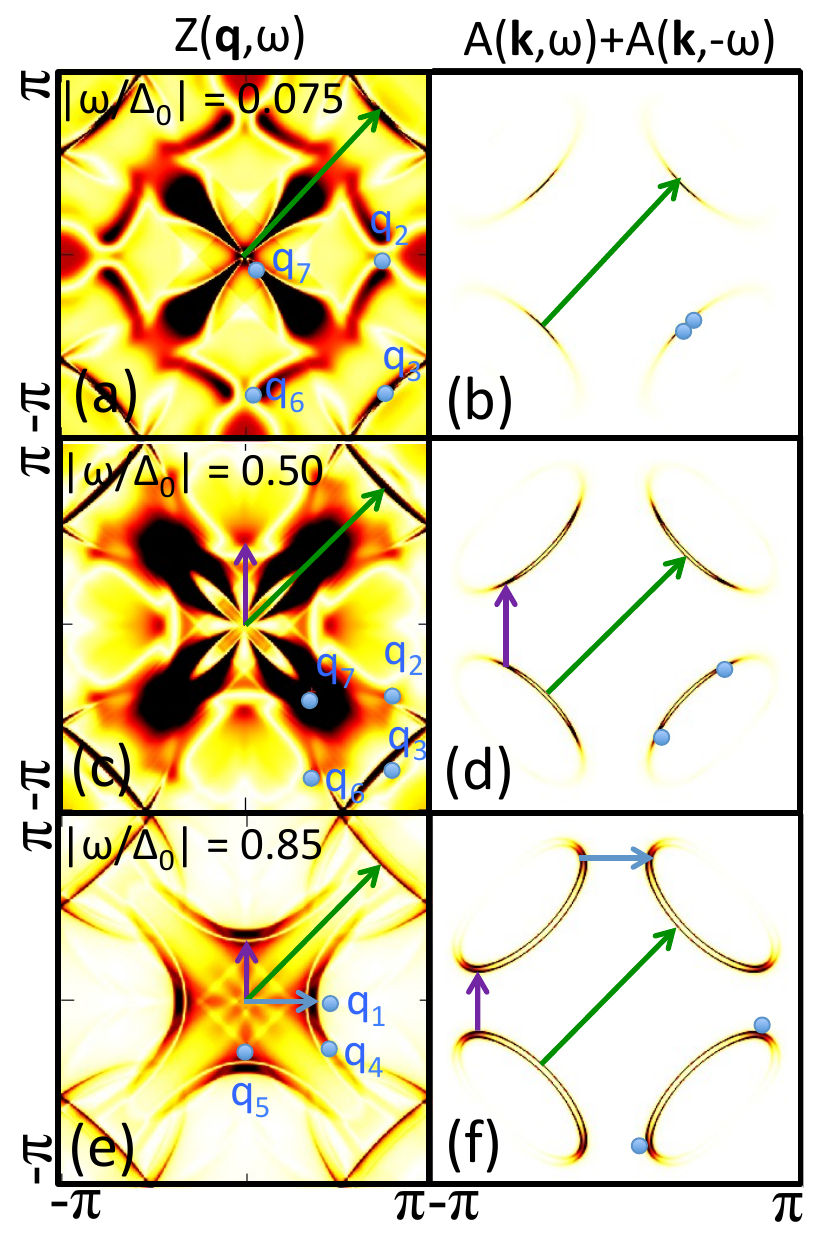}
\end{center}
\caption{\label{fig:dw_and_sc_zmap}  {\bf Coexisting $d$-wave superconductivity and ${\bf Q}=(\pi,\pi)$ long range DW order}. $Z({\bf q},\omega)$ (left column) and sum of spectral function at positive negative bias energies, $A({\bf k},\omega)+A({\bf k},-\omega)$ (right column) for $|\omega/\Delta_{0}|=0.075$ (a and b), $|\omega/\Delta_{0}|=0.5$ (c and d), and $|\omega/\Delta_{0}|=0.85$ (e and f).  Here $\Delta_{0}=35$ meV and $\Delta_{DW}=100$ meV.  Blue circles mark the locations of the octet ${\bf q}$-vectors  on $Z({\bf q},\omega)$, and the ends of the CCE on $A({\bf k},\omega)$, respectively. New features arising from the DW order in $Z({\bf q},\omega)$ are marked with arrows color coded with the corresponding scattering vectors in the spectral function.}
\end{figure}

The FT Z-map is evaluated using the self-consistent $T$-matrix formalism described in Section IV, except the $T$-matrix and Green's function are now $4\times4$. The FT Z-map and the sum of spectral functions at positive and negative bias energies, for coexisting superconductivity and a ${\bf Q}=(\pi,\pi)$ long range DW are shown in Fig.~\ref{fig:dw_and_sc_zmap}.  The DW has an isotropic gap $\Delta_{DW}$=100 meV, while the $d$-wave supeconducting gap maximum is $\Delta_0=$35 meV.  At low energy the DW order has little effect on the spectral function (compare Fig.~\ref{fig:dw_and_sc_zmap}b and Fig.~\ref{fig:zmap-sc}b), because the DW vector does not nest the nodal region. As energy increases and the DW order causes a bend back in the Fermi surface, the spectral function at positive and negative bias energies differs (Fig.~\ref{fig:dw_and_sc_zmap}f).  The ends of the banana-shaped superconducting CCE from Fig.~\ref{fig:zmap-sc} are marked by blue dots, note that the region of maximum spectral intensity now shifts away from these points.  

 Reflecting the energy evolution of the spectral function, at low energy the FT Z-map is fairly similar to the pure superconducting case (compare Fig.~\ref{fig:dw_and_sc_zmap}a and Fig.~\ref{fig:zmap-sc}a), although a strong arc-like feature appears due to scattering between the centers of the CCE (green arrow).  With increasing energy, the FT Z-map deviates more from the superconducting case, so that in Fig.~\ref{fig:dw_and_sc_zmap}c the scattering vectors nesting the ends of the elliptical  CCE show up (purple arrow). Finally, in Fig.~\ref{fig:dw_and_sc_zmap}e, the FT Z-map is dominated by the scattering vectors due to the DW,  discussed in the previous section, rather than the octet ${\bf q}$-vectors. 

\section{Coexistence of superconductivity and fluctuating density wave order}

Despite extensive efforts, the only long range DW orders in the cuprates identified thus far  with bulk sensitive probes are antiferromagnetic spin order in the undoped insulating compounds, and static spin stripe order in doped La$_{(2-x-y)}$(Sr,Ba)$_x$(Nd,Eu)$_y$CuO$_4$.~\cite{tranquada}
 As a result, if the pseudogap arises from a DW order, this order would likely be of  a fluctuating type, which cannot be picked up by these probes.  Motivated by this, we now consider the scenario of superconductivity coexisting with fluctuating DW order.
We again consider two types of DW order: commensurate ${\bf Q}=(\pi,\pi)$ order, and incommensurate checkerboard order with wavevectors ${\bf Q}_2^x=(0.26\pi, 0)$, ${\bf Q}_2^y=(0, 0.26\pi)$.  

The fluctuating DW is included as a self-energy term in the normal state non-interacting Green's function as in Refs.~\onlinecite{Lee, Harrison_afm}:
\begin{equation}
\label{eq:g}
G_0({\bf k},i\omega_n)=\frac{1}{i\omega_n- \epsilon_{\bf k}-\Sigma_{\bf k}(i\omega_n)}
\end{equation}
where the self-energy term is given by 
\begin{equation}
\Sigma_{\bf k}(i\omega_n)=\int d{\bf q} P({\bf q}) \Delta_{DW}^2/(i\omega_n-\epsilon_{{\bf k}+{\bf q}})
\end{equation}
  The function $P({\bf q})$ is a Lorentzian that peaks at the DW ordering vectors. The superconducting Green's function is now given by
\[ \hat{G}^{-1}({\bf k},i\omega_n)= i\omega_n\hat{I}- \left( \begin{array} {cc}
\epsilon_{\bf k}+\Sigma_{\bf k}(i\omega_n) & \Delta_{\bf k} \\
\Delta_{\bf k} & -\epsilon_{\bf k}-\Sigma_{\bf k}(-i\omega_n) \end{array} \right) \] 
This modified Green's function is now inserted into the self-consistent $T$-matrix formalism described in Section IV.

\begin{figure}
\includegraphics[width=0.48\textwidth]{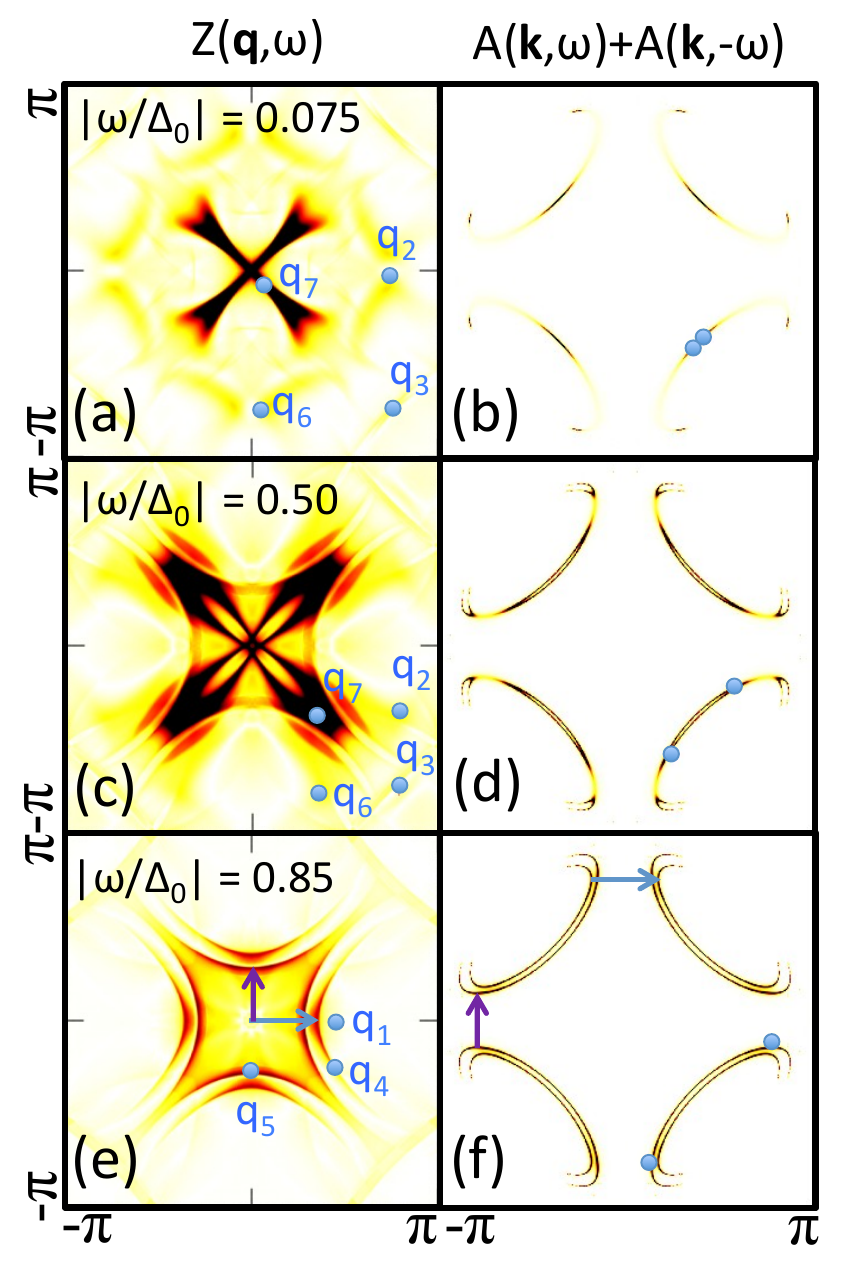}
\caption{\label{fig:fluc_pipi} {\bf Coexisting $d$-wave superconductivity and ${\bf Q}=(\pi,\pi)$ fluctuating DW order}. $Z({\bf q},\omega)$ (left column) and sum of spectral function at positive negative bias energies, $A({\bf k},\omega)+A({\bf k},-\omega)$ (right column) for $|\omega/\Delta_{0}|=0.075$ (a and b), $|\omega/\Delta_{0}|=0.5$ (c and d), and $|\omega/\Delta_{0}|=0.85$ (e and f).  Here $\Delta_{0}=35$ meV and $\Delta_{DW}=100$.  Blue circles mark the locations of the octet  ${\bf q}$-vectors  on $Z({\bf q},\omega)$, and the ends of the CCE on $A({\bf k},\omega)$, respectively. New features arising from the DW order in $Z({\bf q},\omega)$ are marked with arrows color coded with the corresponding scattering vectors in the spectral function. }
\end{figure}

First consider the case of a fluctuating $(\pi, \pi)$ DW coexisting with superconductivity (Fig.~\ref{fig:fluc_pipi}). Compared to long range order (Fig.~\ref{fig:dw_and_sc_zmap}), fluctuating order has less  effect on the FT Z-map, at least at low energy.  The spectral function is similar to the spectral function without DW order (Fig.~\ref{fig:fluc_pipi}b), whereas at higher energies there is a significant Fermi surface bend back and redistribution of spectral weight (Fig.~\ref{fig:fluc_pipi}d, f).  The FT Z-map at low energy (Fig.~\ref{fig:fluc_pipi}a) is nearly identical to the FT Z-map with superconductivity only (Fig.~\ref{fig:zmap-sc}a), except the intensity at the octet $\bf{q}$-vectors is already reduced, reflecting redistribution of spectral weight by the DW.  In Fig.~\ref{fig:fluc_pipi}c, new features start to appear in the FT Z-map, and finally in Fig.~\ref{fig:fluc_pipi}e the ``doubled" peaks from the DW order completely dominate the FT Z-map (purple and blue arrows).  The signatures of fluctuating $(\pi,\pi)$ DW order coexisting with superconductivity are qualitatively similar to those observed for long range order, with superconductivity dominating at low energies and the DW order dominating at larger energies.

\begin{figure*}
\begin{center}
\includegraphics[width=0.72\textwidth]{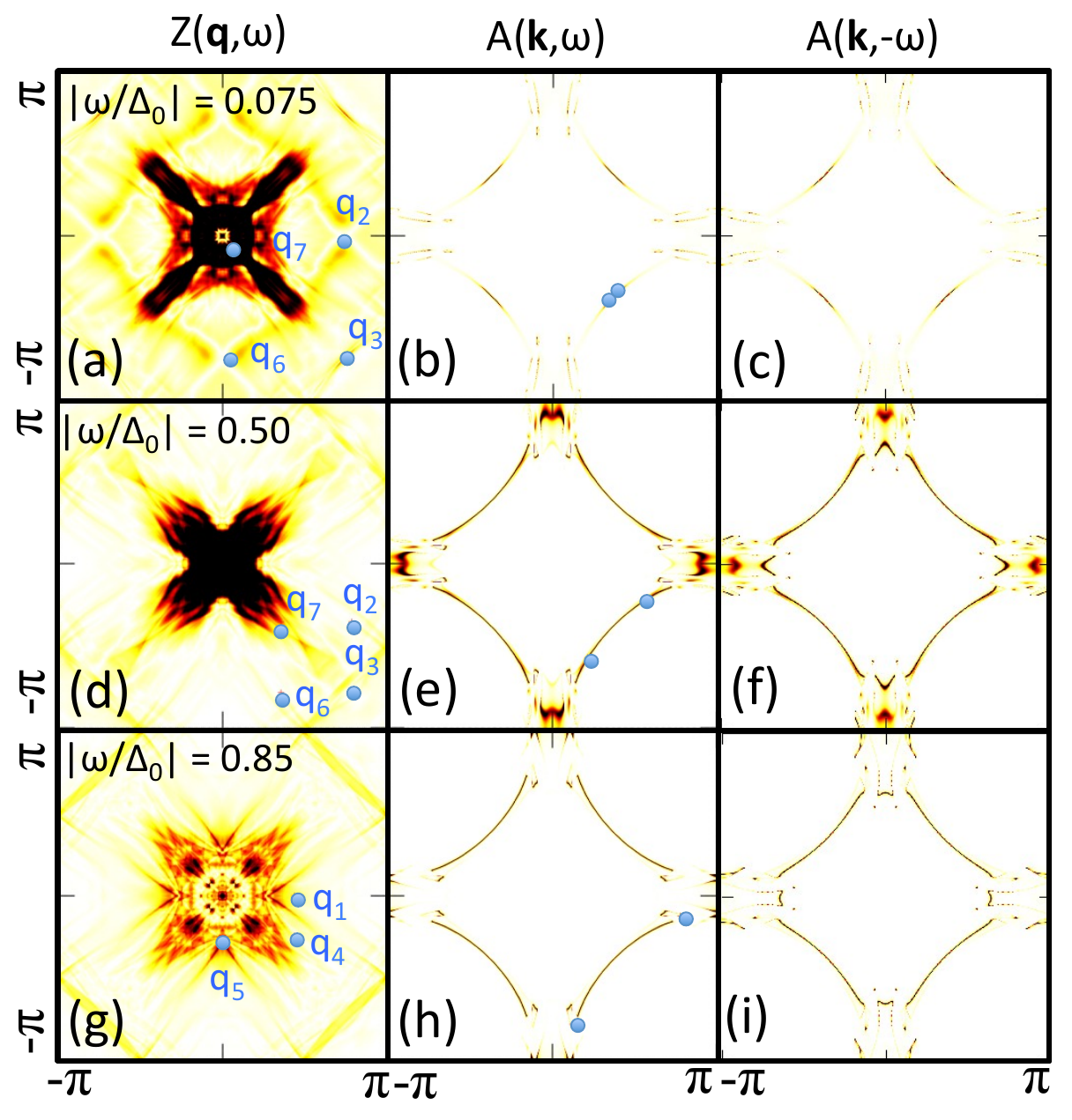}
\end{center}
\caption{\label{fig:cb} {\bf Coexisting $d$-wave superconductivity and  fluctuating checkerboard DW order}.  
$Z({\bf q},\omega)$ (left column), spectral function at positive bias energy $A({\bf k},\omega)$ (middle column), and spectral function at  negative bias energy, $A({\bf k},-\omega)$ (right column) for $|\omega/\Delta_{0}|=0.075$ (a-c), $|\omega/\Delta_{0}|=0.5$ (d-f), and $|\omega/\Delta_{0}|=0.85$ (g-i).  Here $\Delta_{0}=35$ meV and $\Delta_{DW}=100$ meV.  Blue circles mark the locations of the octet ${\bf q}$-vectors  on $Z({\bf q},\omega)$, and the ends of the CCE on $A({\bf k},\omega)$, respectively. New features arising from the DW order in $Z({\bf q},\omega)$ are marked with arrows color coded with the corresponding scattering vector in the spectral function.  The checkerboard wavevectors are set to ${\bf Q}_2^x=(0.26\pi,0)$ and  ${\bf Q}_2^y=(0,0.26\pi)$.}
\end{figure*}

The fluctuating checkerboard order coexisting with superconductivity (Fig.~\ref{fig:cb}) causes a significant breakup of the spectral function at all energies, which is plotted separately for positive and negative bias energy for clarity. Note that the spectral function at $\pm \omega$ looks increasingly different as $|\omega|$ grows. Intensity at the octet ${\bf q}$-vectors is visible in Fig.~\ref{fig:cb}a, however the FT Z-map in Fig.~\ref{fig:cb}d and g is completely dominated by wavevectors introduced by the shifted spectral weight.  Due to the differences between $A({\bf k},\omega)$ and $A({\bf k},-\omega)$, the $Z({\bf q},\omega)$ intensity has separate contributions from each. However,  the large number of scattering vectors obscures  the ``doubling" effect clearly visible for $(\pi,\pi)$ ordering.

To summarize, these results show that, like a long range DW order, a fluctuating DW order has a significant effect on the FT Z-map, and different DW orders produce markedly different results.  Due to particle-hole symmetry breaking in the DW phase, a ``doubling" of peaks occurs in the FT Z-map, which is evident for a simple $(\pi,\pi)$ order.  The Fermi surface is significantly reconstructed in the case of a checkerboard order, introducing a large number of new peaks into the FT Z-map.  As a result, the FT Z-map is a useful tool to both identify the presence of a DW order (for the case of simple orders, where the peak "doubling" is clearly visible), and to distinguish between different types of orders.

\section{Conclusions}
In this paper we  presented a systematic study of the signatures of superconductivity and DW order in the Z-map. For superconducting QPI, we showed how impurities that modulate the bond, superconducting gap, and site energy parameters can be differentiated by the different QPI patterns they produce.  We noted that due to its definition, the Z-map is inherently sensitive to particle-hole asymmetry, and showed that the evolution of its intensity with energy reflects the underlying particle-hole asymmetry of the cuprate bandstructure.

In the second part of the paper, motivated by recent experimental results suggesting a possible DW origin for the pseudogap in the cuprates, we explored the effect of both long range and fluctuating DW orders on the FT Z-map.  The reorganization of the Fermi surface by the DW order introduced new peaks into the FT Z-map. Due to the particle-hole symmetry breaking in a DW phase, scattering wavevectors at positive and negative bias energies are different, leading to a ``doubling" of the wavevectors in the FT Z-map.  For a simple $(\pi,\pi)$ order,  the features can be easily connected to scattering vectors between regions of large intensity in the spectral function.  For the more complicated checkerboard order, this is not possible.  However, we note that different types of DW order produce very different  signatures, providing a means of distinguishing between them.  

As is evident from this discussion, the Z-map contains a great deal of information. However, due to this complexity, features due to different types of impurity scattering and particle-hole asymmetries of various origins may obscure the signals from underlying DW orders.  In addition, STS resolution issues and tunneling matrix elements, not considered here, may complicate  detection of the subtle differences in the Z-map patterns from superconductivity and different types of DW order. This could explain why the signatures  discussed in this paper have not been detected, even if present in the system.

Finally, we note that in the phase diagram of many materials, there is a proximity between DW order and superconductivity, for example, the transition from antiferromagnetic spin order to superconductivity as a function of doping in both the cuprate and pnictide families of superconductors, coexisting CDW order and conventional superconductivity in NbSe$_2$,~\cite{borisenko} and the pressure-induced transition from a CDW to a superconducting state in the rare earth tritelluride compounds ($R$Te$_3$).~\cite{rte3}
As a quantity that can distinguish between these phases, the Z-map may be useful in unravelling the relation between superconductivity and charge and spin orders in many compounds.

\section{Acknowledgements}
We thank M. Hashimoto, R.-H. He, P. J. Hirschfeld, J. E. Hoffman, A. F. Kemper, and I. M. Vishik for useful discussions. We acknowledge support 
from  the U. S. Department of Energy, 
Office of Basic Energy Science, Division of Materials Science and 
Engineering under Contract No. DE-AC02-76SF00515.

\newpage

\bibliographystyle{apsrev}
\bibliography{nowadnick}

\end{document}